\documentclass[a4paper]{jpconf}
\usepackage{graphicx}
\usepackage{hyperref}
\usepackage{amsmath}
\usepackage{physics}
\usepackage[english]{babel}
\usepackage[utf8]{inputenc}

\newcommand{\PowR}{\mathcal{P}_\mathcal{R}}
\newcommand{\R}{\mathcal{R}}
\newcommand{\phil}{\bar\phi}
\newcommand{\pil}{\bar\pi}

\begin{document}
\title{A numerical approach to stochastic inflation and primordial black holes}

\author{Eemeli Tomberg}

\address{Laboratory of High Energy and Computational Physics, National Institute of Chemical Physics and Biophysics, R\"{a}vala puiestee 10, 10143 Tallinn, Estonia}

\ead{eemeli.tomberg@kbfi.ee}

\begin{abstract}
Certain models of cosmic inflation produce strong cosmological perturbations at short length scales, which may later collapse into primordial black holes. To find the statistics of these strong perturbations and the ensuing black holes, it is necessary to go beyond linear perturbation theory. Stochastic inflation provides a way to take the leading non-linear effects into account. In this contribution, I discuss recent progress in numerical computations of stochastic inflation. A numerical approach can include more of the non-linearities than an analytical calculation, and can be applied to single-field inflationary models with any potential.
\end{abstract}

\section{Introduction}
\label{sec:intro}
Primordial black holes (PBHs) are hypothetical objects produced in many models of early universe cosmology. If observed today, they would carry information on the conditions in these early times. PBHs were first discussed in the 1970s \cite{Carr:1974nx}, but with improving observational techniques---in particular, the first detection of gravitational waves from two merging black holes \cite{LIGOScientific:2016aoc}---their study has been revitalized over the recent years.

One way to produce PBHs is to first produce strong scalar perturbations during cosmic inflation, which later collapse into black holes. The simplest, canonical single-field models of inflation are fully described by the scalar field potential. The task is then to relate each potential to an ensuing abundance and distribution of black holes. This task is complicated due to the extreme nature of the black holes: to not overclose the universe, they must only form in small numbers from the strongest perturbations, which tend to lie beyond the scope of usual linear perturbation theory. In this contribution, I describe recent progress in computing these strong perturbations with the tools of stochastic inflation in a numerical setup, as first reported in \cite{Figueroa:2020jkf}.

\section{Primordial black holes}
\label{sec:PBHs}
Primordial black holes are born in the early universe, typically during the radiation dominated era and before the Big Bang nucleosynthesis. Later, they may contribute to the dark matter, or act as seeds of the supermassive black holes in the centers of galaxies. Observations from gravitational lensing and other sources have constrained the allowed PBH mass and abundance, but there are still windows in which PBHs can contribute all of the dark matter, especially in the range $10^{17}$--$10^{22}$ g \cite{Carr:2020xqk, Green:2020jor}.

A possible formation mechanism for PBHs are strong primordial scalar perturbations seeded during inflation. These perturbations originate from quantum vacuum fluctuations and are also responsible for the anisotropies seen in the cosmic microwave background (CMB) radiation. Perturbations on these long scales are weak, with a power spectrum of the comoving curvature perturbation $\PowR \sim 10^{-9}$ \cite{Aghanim:2018eyx}; PBHs of a reasonable mass would form from perturbations of a much shorter wavelength and would need to be stronger, maybe $\PowR \sim 10^{-2}$, to have a non-negligible abundance. To achieve this enhancement of power, the inflaton potential typically has a feature resembling a saddle point or a local, shallow minimum. As the inflaton rolls over the feature, its dynamics follow the so-called ultra-slow-roll inflation, associated with enhanced perturbations. I discuss below an example scenario like this, originally studied in \cite{Figueroa:2020jkf}, with the potential depicted in figure~\ref{fig:pot}. The feature in the potential has been tuned carefully to not overproduce or underproduce PBHs.

The perturbations created during inflation quickly get stretched to super-Hubble scales, where they stay frozen. The expansion of space stops even strong perturbations from collapsing gravitationally. However, after inflation, the expansion decelerates and the perturbation scales start to re-enter the Hubble radius. When this happens, all matter in a patch of roughly the Hubble size collapses into a black hole if the local perturbations are strong enough. This gives a correspondence between the mass of the black holes and the length scale of the perturbations, or equivalently the era of inflation the perturbations originated from. In practice, the largest black holes originate from earlier times during inflation but collapse later during radiation domination. In the example case, I consider PBHs from the strong perturbations that exited the Hubble radius at the end of the ultra-slow-roll region. The corresponding mass is $1.4 \times 10^{19}$ g, in the observationally allowed window mentioned above.

In practice, the strength of the perturbations is measured by the comoving curvature perturbation $\R$. If $\R$, averaged (`coarse-grained') over a patch of scale $k_c$, exceeds a collapse threshold of roughly $\R_\mathrm{th} \approx 1$ in a patch, then this patch becomes a PBH when $k_c$ re-enters the Hubble radius. (This description of the collapse is subject to many caveats related to, for example, the window function used in coarse-graining, the relationship between $\R$ and the local density perturbation, and the shape of the collapsing perturbations; I omit such complications to keep the discussion simple.) The statistics of the PBHs then follow the statistics of the coarse-grained curvature perturbation and can be obtained by the methods of stochastic inflation.

\section{Stochastic inflation}
\label{sec:stoch}
Stochastic inflation is a method for computing cosmological perturbations from inflation beyond linear perturbation theory. The method was first discussed in \cite{Starobinsky:1986fx}; for a recent, modern treatment, see e.g. \cite{Pattison:2017mbe, Pattison:2019hef}.
In this method, the inflaton field is divided into long and short wavelength modes:
\begin{equation}
    \phi(t,\vec{x}) =
    \int\limits_{k<k_c} \frac{\rmd^3 k}{(2\pi)^\frac{3}{2}} \phi_{\vec{k}}(t) e^{-i\vec{k} \cdot \vec{x}} +
    \int\limits_{k>k_c} \frac{\rmd^3 k}{(2\pi)^\frac{3}{2}} \phi_{\vec{k}}(t) e^{-i\vec{k} \cdot \vec{x}}
    \equiv \phil(t,\vec{x}) + \delta\phi(t,\vec{x}) \, ,
\end{equation}
divided by the coarse-graining scale $k_c=\sigma aH$, where $\sigma \ll 1$ is a constant. The coarse-grained long-wavelength field $\phil$ only varies over length scales longer than $1/k_c$; in one patch of size $1/k_c$, it is roughly constant in space. Since such a patch is much larger than the Hubble radius ($\sigma \ll 1$), the field value $\phil$ there follows local FLRW equations, up to subleading gradient terms. Fast spatial expansion keeps different patches separated, and they evolve independently of each other.

However, the short-wavelength perturbations affect the evolution of the `local background' given by $\phil$. As space expands, the short-wavelength Fourier modes get stretched until they eventually exceed the coarse-graining scale and join the coarse-grained field $\phil$. This introduces random kicks to the evolution of $\phil$ and its time derivative, denoted below by $\pil$, random due to the quantum nature of the short-wavelength perturbations. This manifests as stochastic noise in the equations of motion.

To be more quantitative, the equations solved for each patch in the setup of \cite{Figueroa:2020jkf} are
\begin{equation}
    \label{eq:bg_eoms}
	\phil' = \pil + \xi_\phi \, ,
	\qquad
	\pil' = - \qty( 3 + \frac{H'}{H} ) \pil - \frac{V_{,\phil}}{H^2} + \xi_\pi \, ,
	\qquad
	2 V = ( 6 - \bar\pi^2 ) H^2
\end{equation}
for the local background, with $V$ the inflaton potential and $H$ the Hubble parameter, and
\begin{equation}
    \label{eq:perts_eom}
    \delta\phi_k'' + \left(3  + \frac{H'}{ H}\right) \delta\phi'_k + \omega_k^2 \delta\phi_k = 0 \, , \quad \omega_k^2 \equiv \frac{k^2}{(aH)^2} +\pil^2 \qty(3 + 2\frac{H'}{H} - \frac{H'}{H^2}) + 2\pil \frac{V_{,\phil}}{H^2} + \frac{V_{,\phil\phil}}{H^2}
\end{equation}
for the short-wavelength perturbations.  The time variable is the number of e-folds of spatial expansion $N$, and derivatives with respect to it are denoted by a prime. This choice of a time variable is customary and discussed further below. The standard FLRW evolution of $\phil$ and $\pil$ is augmented by the stochastic noise terms $\xi_\phi$ and $\xi_\pi$, which depend on the short-wavelength perturbations, treated here at the linear level. The short-wavelength perturbations are treated quantum mechanically; the mode functions $\delta\phi_k$ contain the time dependence of the corresponding quantum operators, and they start from the usual Bunch-Davies vacuum values deep inside the Hubble radius. As the modes stretch, they become `squeezed' and their probability distribution starts to resemble that of classical perturbations. The ensuing classical, stochastic noise to the local background is white and Gaussian, with the two-point function \cite{Pattison:2019hef}
\begin{equation}
    \label{eq:xi_exp}
    \expval{\xi_\phi(N) \xi_\phi(N')} = \frac{1}{6\pi^2} \frac{\rmd (\sigma aH)^3}{\rmd N} |\delta\phi_{k=\sigma aH}|^2 \delta(N-N')
\end{equation}
and with the noises correlated as $\xi_\pi = \xi_\phi \frac{\phi'_k}{\phi_k}$. In most studies of stochastic inflation, the noise is taken to follow the zeroth-order result from slow-roll inflation, so that $\expval{\xi_\phi(N) \xi_\phi(N')} = \qty(\frac{H}{2\pi})^2 \delta(N-N')$, and the equations are solved semi-analytically. The study \cite{Figueroa:2020jkf} improves upon this in two important ways: the mode functions $\delta\phi_k$ are solved explicitly from \eqref{eq:perts_eom}, and furthermore, their evolution depends on the local background $\phil$, creating a feedback loop with backreaction effects between the local background and the noise. Such non-Markovian effects can only be captured by solving the system \eqref{eq:bg_eoms}--\eqref{eq:perts_eom} numerically, patch by patch. The inclusion of backreaction is one way in which the stochastic formalism improves on usual linear perturbation theory, where the Fourier modes evolve independently in a global background; an additional and even more important source of non-linearity comes from the FLRW-like equations \eqref{eq:bg_eoms}, in particular the non-linear potential derivative. The hope is that this captures the most important interactions between different scales. The stochastic treatment is essentially a middle ground between a simpler, linear computation and a full numerical simulation of general relativity. The latter, while more accurate, would be hopelessly slow, especially since it would need to cover a large volume to gather statistics for the rare strong perturbations that form PBHs.

To complete the picture, the stochastic computation needs to be connected to cosmological perturbation theory and the comoving curvature perturbation $\R$. This is done by the $\Delta N$ formalism, which tells that the coarse-grained $\R$ of a patch is equal to $\Delta N \equiv N - \bar{N}$, where $N$ is the number of e-folds of local expansion between an initial unperturbed hypersurface with a fixed $\phil = \phil_i$ and a final hypersurface with $\phil = \phil_f$, and $\bar{N}$ is the distribution average \cite{Sasaki:1995aw}. This is the reason $N$ was chosen as the time variable in \eqref{eq:bg_eoms}--\eqref{eq:perts_eom}: now $N$ is simply the duration of the simulation; in particular, it does not receive stochastic kicks. There are additional subtleties related to the time variable and the choice of the gauge of cosmological perturbations; see \cite{Figueroa:2020jkf} for further discussion.

\begin{figure}[t]
\begin{minipage}{17pc}
\includegraphics[scale=0.8]{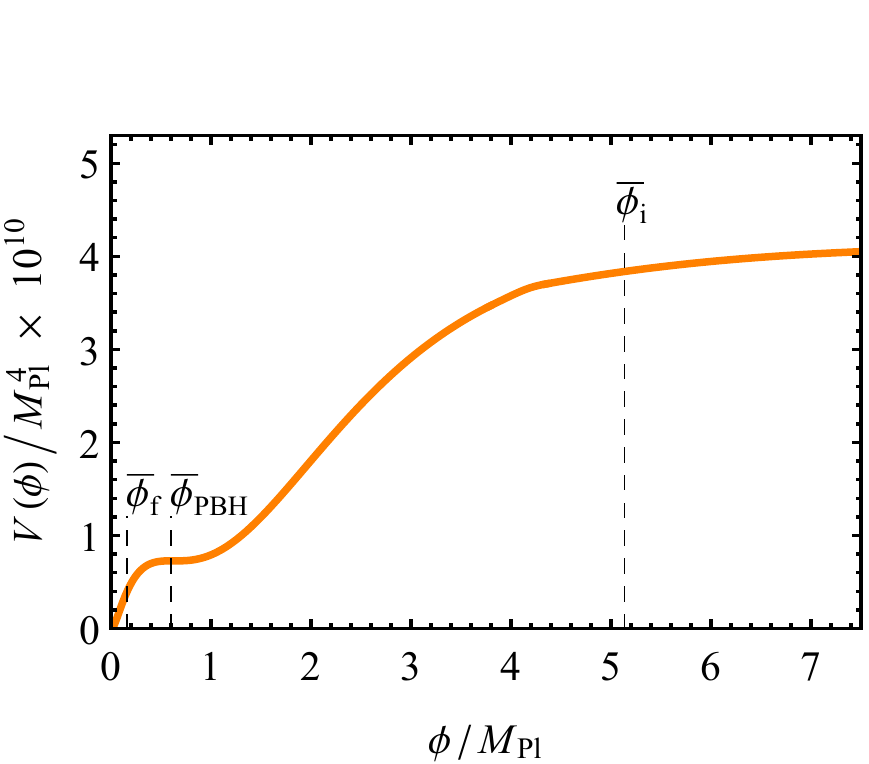}
\caption{\label{fig:pot}The potential considered in \cite{Figueroa:2020jkf}, adapted from figure~1 therein, compatible with the CMB measurements \cite{Aghanim:2018eyx}.}
\end{minipage}\hspace{4pc}%
\begin{minipage}{17pc}
\hspace{-2pc}
\includegraphics[scale=0.8]{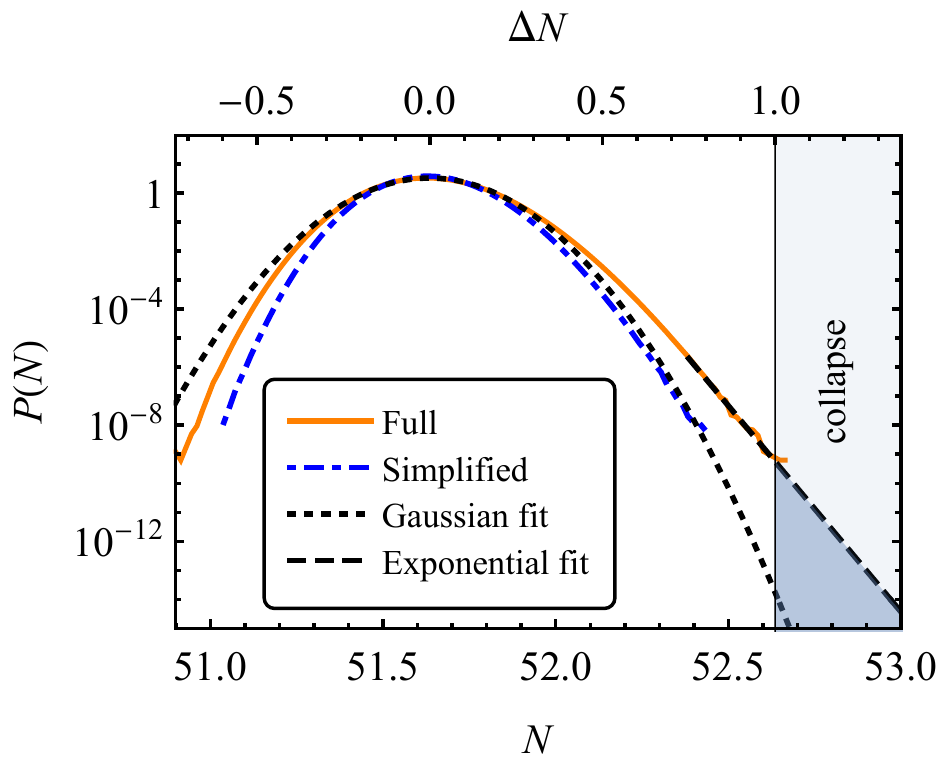}
\caption{\label{fig:R_distr}Probability distribution of the local amount of expansion in a patch, adapted from figure~2 of \cite{Figueroa:2020jkf}.}
\end{minipage} 
\end{figure}

The algorithm for solving the PBH statistics numerically is then as follows:
\begin{itemize}
    \item Follow one patch of super-Hubble size, starting from the CMB scale $\phil_i$ for convenience, with perturbations initially in their vacuum state.
    \item Evolve the coupled system \eqref{eq:bg_eoms}--\eqref{eq:perts_eom}, pulling random stochastic kicks from the probability distribution of the noise given by \eqref{eq:xi_exp} at each time step.
    \item Stop the kicks when the desired PBH scale $k_\mathrm{PBH}$ is reached (roughly at $\phil_\mathrm{PBH}$ in figure~\ref{fig:pot}).
    \item Run the local background evolution without kicks to a final hypersurface with $\phil_f$ near the end of inflation.
    \item Record the final time variable $N$. Repeat and build statistics to find the probability distribution of $\Delta N = \R$.
\end{itemize}
Subtleties related to the discretization of the continuum equations and the numerical implementation are discussed further in \cite{Figueroa:2020jkf}. An important point is stopping the kicks when $k_c = k_\mathrm{PBH}$, a pre-fixed PBH scale. This algorithm gives the distribution of $\R$ coarse-grained over the fixed scale $k_\mathrm{PBH}$, thus related to PBHs of a fixed mass as discussed in section~\ref{sec:PBHs}. Originally, while the coarse-graining scale $k_c=\sigma aH$, it evolves in time, moving to shorter and shorter comoving length scales as the universe expands. Once $k_c$ has reached the value $k_\mathrm{PBH}$, we want this evolution to stop, so we freeze $k_c$ to this value. No modes exit the coarse-graining scale after this, giving no more kicks. Shorter-wavelength perturbations are `averaged over' and don't affect the statistics of PBHs of this mass scale. If one wants to study the full PBH distribution over all mass scales, the simulations need to be rerun with different values of $k_\mathrm{PBH}$.

\section{Numerical results}
\label{sec:results}
In \cite{Figueroa:2020jkf}, $10^{11}$ simulations were run for the chosen PBH mass scale $1.4 \times 10^{19}$ g to obtain the $\Delta N = \R$ distribution of figure~\ref{fig:R_distr}. The initial PBH abundance is obtained by integrating the distribution over the tail where $\R > \R_\mathrm{th}=1$. An abundance of roughly $10^{-16}$ is needed to match the observed dark matter density today. The potential is chosen so that a Gaussian fit, marked by a dotted line in the figure and valid close to the mean, matches this requirement. The Gaussian fit matches a linear perturbation theory estimate which is much easier to compute than the full numerical result, and for this reason it was used to fine-tune the potential. Because of their ease of computation, Gaussian approximations are common in PBH studies in general. However, the fit is not good for large $\R$; instead, the full distribution follows an exponential behavior with $P(\R) \propto e^{-33\R}$ in the tail. Such exponential tails have been discussed before in the context of semi-analytical toy models in \cite{Pattison:2017mbe, Ezquiaga:2019ftu}; in \cite{Figueroa:2020jkf}, they were shown to exist also for a realistic model which fits the CMB observations.

The exponential tail enhances the PBH abundance considerably, here by a factor of $10^5$. (This corresponds to an initial abundance of $10^{-11}$, roughly the inverse of the number of simulations, $10^{11}$, needed to probe the tail at this depth.) The abundance from the full computation is also enhanced with respect to the simplified computation using the leading-order noise discussed below \eqref{eq:xi_exp}. This shows the importance of accounting for the non-linearities correctly when estimating the PBH abundance.

\section{Conclusions}
\label{sec:conclusions}
I have explained how stochastic inflation can be used to numerically compute the abundance of PBHs formed from primordial perturbations. The method requires a lot of computational resources, but is more accurate than previous, semi-analytical methods, and includes backreaction effects between the background and the stochastic noise. An example computation confirms that the perturbation distributions have exponential tails, which enhance the PBH abundance considerably compared to traditional Gaussian estimates. The algorithm presented here can be applied to any single-field inflationary potential.

Paper \cite{Figueroa:2020jkf} should be seen as a proof of concept for these types of stochastic inflation computations. Further work by the authors is underway, including  more numerical examples showing the generality of the exponential tails. In the future, the full mass spectrum of PBHs, together with correlations between different scales, can be studied with similar methods. There are also unresolved issues related to e.g. the fundamental derivation of the stochastic formalism in ultra-slow-roll inflation and the choice of the coarse-graining scale; see \cite{Figueroa:2020jkf} for further discussion.

%I have explained how stoch infl can be used to compute distr of primordial cosm perts, and discussed a more-accurate-than-before numerical way to do this, which includes more non-linearities and backreaction. Needed to compute ensuing PBH abundance.

%We see an exp tail in pert dist. Effects like this increase abundance, (may make fine-tuning a LITTLE less severe, thouhg not a lot), affect staistics, may affect ini distribution. Correlations? Work to do ("A prototypal investigation: in the future, similar methods can be used to study..."). Also, ambiguities related to stoch formalism in USR, choice of coarse-gr scale, gauge issues, etc that we didn't go into details of, see (cite us) and references therein.

%Can repeat for any potential.

\ack
This work was supported by the Estonian Research Council grants PRG803, PRG1055, and MOBTT5 and by the EU through the European Regional Development Fund CoE program TK133 ``The Dark Side of the Universe''. The work in \cite{Figueroa:2020jkf} was done in collaboration with Daniel Figueroa, Sami Raatikainen, and Syksy R\"{a}s\"{a}nen.

%%%%%%%%%%%%

% ?From future work: comments: non-Marovianities not important. Similar stuff seen for many potentials.

\section*{References}

\bibliography{bibliography.bib}

\providecommand{\newblock}{}
\begin{thebibliography}{10}
\expandafter\ifx\csname url\endcsname\relax
  \def\url#1{{\tt #1}}\fi
\expandafter\ifx\csname urlprefix\endcsname\relax\def\urlprefix{URL }\fi
\providecommand{\eprint}[2][]{\url{#2}}
% Bibliography created with iopart-num v2.0
% /biblio/bibtex/contrib/iopart-num

\bibitem{Carr:1974nx}
Carr B~J and Hawking S~W 1974 {\em Mon. Not. Roy. Astron. Soc.\/} {\bf 168}
  399--415

\bibitem{LIGOScientific:2016aoc}
Abbott B~P {\em et~al.\/} (LIGO Scientific, Virgo) 2016 {\em Phys. Rev.
  Lett.\/} {\bf 116} 061102 (\textit{Preprint} \eprint{1602.03837})

\bibitem{Figueroa:2020jkf}
Figueroa D~G, Raatikainen S, Rasanen S and Tomberg E 2021 {\em Phys. Rev.
  Lett.\/} {\bf 127} 101302 (\textit{Preprint} \eprint{2012.06551})

\bibitem{Carr:2020xqk}
Carr B and Kuhnel F 2020 {\em Ann. Rev. Nucl. Part. Sci.\/} {\bf 70} 355--394
  (\textit{Preprint} \eprint{2006.02838})

\bibitem{Green:2020jor}
Green A~M and Kavanagh B~J 2021 {\em J. Phys. G\/} {\bf 48} 043001
  (\textit{Preprint} \eprint{2007.10722})

\bibitem{Aghanim:2018eyx}
Aghanim N {\em et~al.\/} (Planck) 2020 {\em Astron. Astrophys.\/} {\bf 641} A6
  [Erratum: Astron.Astrophys. 652, C4 (2021)] (\textit{Preprint}
  \eprint{1807.06209})

\bibitem{Starobinsky:1986fx}
Starobinsky A~A 1986 {\em Lect. Notes Phys.\/} {\bf 246} 107--126

\bibitem{Pattison:2017mbe}
Pattison C, Vennin V, Assadullahi H and Wands D 2017 {\em JCAP\/} {\bf 10} 046
  (\textit{Preprint} \eprint{1707.00537})

\bibitem{Pattison:2019hef}
Pattison C, Vennin V, Assadullahi H and Wands D 2019 {\em JCAP\/} {\bf 07} 031
  (\textit{Preprint} \eprint{1905.06300})

\bibitem{Sasaki:1995aw}
Sasaki M and Stewart E~D 1996 {\em Prog. Theor. Phys.\/} {\bf 95} 71--78
  (\textit{Preprint} \eprint{astro-ph/9507001})

\bibitem{Ezquiaga:2019ftu}
Ezquiaga J~M, Garc\'\i{}a-Bellido J and Vennin V 2020 {\em JCAP\/} {\bf 03} 029
  (\textit{Preprint} \eprint{1912.05399})

\end{thebibliography}

\end{document}